\documentstyle[aps,prl,epsf,floats]{revtex} 
\begin{document}

\title{\bf{ Neutrino Models of Dark Energy}}

\author{R. D. Peccei\\}
\address{Department of Physics and Astronomy\\
University of California at Los Angeles\\
Los Angeles, California, 90095}
\maketitle
\vspace{1cm}

\begin{abstract}

I consider a scenario proposed by Fardon, Nelson and Weiner where dark energy and neutrinos are connected. As a result, neutrino masses are not constant but depend on the neutrino number density. By examining the full equation of state for the dark sector, I show that in this scenario the dark energy is equivalent to having a cosmological constant, but one that "runs" as the neutrino mass changes with temperature. Two examples are examined that illustrate the principal features of the dark sector of this scenario. In particular, the cosmological constant is seen to be negligible for most of the evolution of the Universe, becoming inportant only when neutrinos become non-relativistic. Some speculations on features of this scenario which might be present in a more realistic theory are also presented.
\end{abstract}

\section{Introduction}

One of the truly challenging and open questions in both cosmology and
particle physics is the nature of the dark energy in the Universe.
Indeed, I would argue that it is of fundamental importance for both
fields to try to understand the {\bf connection} of the dark energy in
the Universe with particle physics. In this paper I would like to pursue
the consequences of an idea proposed recently by Fardon, Nelson and
Weiner (FNW) \cite{FNW} which tries to tie together the dark energy sector
with that of neutrinos. Although this is a rather speculative idea, it
leads to some interesting results whose implications are very much
worthwhile exploring.

To set the stage, it is useful to write Einstein's equations describing
the expansion of the Universe in a Robertson-Walker background as an
equation for the Hubble parameter $H$ 
\begin{equation}
H^2\equiv (\frac {\dot{R}}{R})^2= \frac{ 8 \pi G_N \rho}{3}, 
\end{equation}
and its rate of change
\begin{equation}
\frac{\ddot{R}}{R} =  -\frac{4 \pi G_N}{3}(\rho +3p)
\end{equation}
In the above, $\rho$  and $p$ are, respectively, the {\bf total} energy density and pressure of the Universe. The dominance of a dark energy component with negative pressure in the present era is responsible for the
Universe's accelerated expansion. 

If the dark energy were due to
the presence of a cosmological constant $\Lambda$, one would have
\begin{equation}
\Lambda =  4 \pi G_N\rho_{\rm {dark~ energy}} = -4 \pi G_Np_{\rm {dark~
energy}},
\end{equation}
corresponding to a dark energy equation of state
\begin{equation}
\omega_{\rm {dark~ energy}} =\frac{p_{\rm {dark~ energy}}}{\rho_{\rm
{dark~ energy}}}= -1.
\end{equation}

Because the Hubble parameter now $ H_o = (1.5 \pm 0.1) 10^{-33}$ eV is a
tiny scale, the intrinsic energy scale associated with dark energy is
also very small. For example, if the dark energy were due to the
presence of a cosmological constant, so that $\rho_{\rm{dark ~energy}}=
\rho_{\rm{vacuum}}= E_o^4$, one finds that $E_o \simeq 2 \times 10^{-3}$
eV. What physics is associated with this small energy scale? All vacuum
energies we know in particle physics are enormously bigger. For
instance, $E_o^{QCD} \sim \Lambda_{QCD} \sim 1$ GeV. It is clearly a
challenge to understand dynamically how the small energy scale
associated with the dark energy density arises and how it is connected to
particle physics.

This is not, however, the only puzzling feature of dark energy. In
parallel, one would like to explain also why, in the present epoch, the
energy density associated with dark energy and that of matter should be
approximately the same: $\rho_{\rm{dark ~energy}} \simeq
\rho_{\rm{matter}}$. Because these densities have different temperature
dependences, their near equality now itself is a mystery. As many people
have noted, this coincidence is resolved dynamically if the dark energy
density tracks (some component) of the matter density. \cite{tracking}

What Fardon, Nelson and Weiner \cite{FNW} suggested is that
$\rho_{\rm{dark~ energy}}$  tracks the energy density in neutrinos,
$\rho_{\nu}$. This avoids some of the issues that would arise if
$\rho_{\rm{dark ~energy}}$ were really to track some better known
component of $\rho_{\rm{matter}}$, like $\rho_{B}$.  Furthermore, it perhaps
allows one to understand the scale $E_o$ associated with
$\rho_{\rm{dark~ energy}}$ in terms of the scale of neutrino masses,
which are of a similar magnitude. However, the idea put forth by Fardon,
Nelson and Weiner is quite radical. By imagining that the
neutrinos and the dark energy are coupled together, the energy density
associated with dark energy depends on the neutrino masses. In turn,
these masses are not fixed but are variable, with their magnitude being
a function of the neutrino density
$m_{\nu}=m_{\nu}(n_{\nu})$.\cite{variable} \cite{JGB} \cite{PQH}

\section{The FNW Scenario}

Fardon {\it et al} \cite{FNW} consider for simplicity just one flavor of
neutrinos. Although it is straightforward to generalize their
considerations to three families of neutrinos, given the exploratory
nature of these considerations, in what follows I will restrict
myself to this simplified case. Indeed, in my opinion, the FNW scenario, on the main, should be thought more as a testing ground for ideas on how to connect dark energy to particle physics than as a truly realistic scenario for dark energy. Hence, for example, in a more realistic theory the dynamical dependence of the neutrino mass with temperature may well be replaced by having all mass parameters dynamically depend on temperature as the Universe evolves.

The fundamental assumption that is
made in the FNW picture is that the energy density in the dark sector
has {\bf two components}. In addition to some form of dark energy (think
of quintessence \cite{Q} \cite{q} \cite{qQ} as an example) the dark sector energy density
also includes the energy density of neutrinos. Thus,
\begin{equation}
\rho_{{\rm dark}}=\rho_{\nu} +\rho_{\rm{dark~ energy}}.
\end{equation}
The neutrinos and the dark energy are coupled because it is assumed that
the dark energy density is a function of the mass of the neutrinos;
$\rho_{\rm{dark~ energy}}= \rho_{\rm{dark~ energy}}(m_{\nu})$. Since
in the present epoch neutrinos are
 non-relativistic, effectively  one can take $\rho_{\nu}=
m_{\nu}n_{\nu}$ and one has
\begin{equation}
\rho_{{\rm dark}}=m_{\nu}n_{\nu} +\rho_{\rm{dark ~energy}}(m_{\nu}).
\end{equation}

Fardon, Nelson and Weiner \cite{FNW} considered the consequences of the
above equation assuming that $\rho_{{\rm dark}}$ is {\bf stationary}
with respect to variations in the neutrino mass. Effectively, this makes
the neutrino mass a variable parameter that changes as the Universe
evolves in time.  This is probably the most far-reaching assumption of FNW. Neutrino masses, in their view, are fixed by the competition between the neutrino energy density $\rho_\nu$ and the energy density of dark energy $\rho_{\rm{dark~energy}}$. While $\rho_\nu$ at high temperatures grows as $T^4$, essentially independent of the value of the neutrino mass, $\rho_{\rm{dark~energy}}$ grows less rapidly at high temperatures and becomes important only when neutrinos become non-relativistic. Stationarity of $\rho_{{\rm dark}}$ under changes in
$m_{\nu}$ implies that
\begin{equation}
\frac{\partial \rho_{\rm{dark}}}{\partial m_{\nu}}=n_{\nu} +\frac{\partial\rho_{\rm{dark ~energy}}(m_{\nu})}{\partial m_{\nu}} =0.
\end{equation}
Given a particular dark energy model, one can use this equation to infer
the dependence of $m_{\nu}$ on the Universe's scale parameter $R$, or its
temperature $T$.

FNW use Eq. (7) and the conservation of energy equation in the
Robertson Walker background (which follows from Einstein's equations)
\begin{equation}
\dot{\rho}=- 3H(\rho + p)
\end{equation}
to deduce the equation of state in the dark sector. Because I will explicitly
derive the equation of state for neutrinos of arbitrary velocities in the next Section, I
will not enter into details here, but just quote the final result.
Defining $\omega$ as
\begin{equation}
\omega = \frac{p_{dark }}{\rho_{dark}}
\end{equation}
one can show that in the FNW scenario $\omega$ obeys the equation of
state
\begin{equation}
\omega +1= \frac {m_{\nu}n_{\nu}}{\rho_{\rm{dark}}}= \frac{m_{\nu}n_{\nu}}{
m_{\nu}n_{\nu}+\rho_{\rm{dark ~energy}}}.
\end{equation}

We see from this equation that if $\omega \simeq -1$, as cosmological
data imply, \cite{Omega} then the neutrino contribution to
$\rho_{\rm{dark}}$ is a small fraction of $\rho_{\rm{dark}}$.
Furthermore, if $\omega$ does not change significantly with $R$, then
both $\rho_{\nu}$ and  $\rho_{\rm{dark}}$ must have the same
dependence on $R$.  From Eq. (8) it is easy to deduce that
$\rho_{\rm{dark}} \sim R^{- 3(1+\omega)}$. Because the neutrino
number density density scales as $n_{\nu} \sim R^{-3}$, it follows that the neutrino
mass is nearly {\bf inversely proportional} to the neutrino density:
\begin{equation}
m_{\nu} \sim n_{\nu}^{\omega}\simeq 1/n_{\nu}.
\end{equation}

\section{A Closer Look at the FNW Scenario}

I have studied the FNW scenario further by examining this scenario
also when neutrinos have arbitrary velocities. As will be seen below, it turns out that these more general considerations
have important consequences also for the non-relativistic limit consider
by Fardon {\it et al}. In the more general case, the energy density for
one generation of neutrinos and antineutrinos is given by
\begin{equation}
\rho_{\nu}= T^4F(\xi),
\end{equation}
where $\xi=m_{\nu}(T)/T$ and \footnote{More precisely, the Fermi factor in Eq. (13) should read [$exp(\sqrt{\xi_d^2 +y^2}) +1$], with $\xi_d=m_{\nu}(T_d)/T_d$ characterizing the neutrino number density after they decoupled at a temperature $T_d$. However, for all practical purposes, $\xi_d$ is totally negligible.}
\begin{equation}
F(\xi)=\frac{1}{\pi^2}\int^{\infty}_0\frac{dyy^2\sqrt{y^2+\xi^2}}{e^y+
1}.
\end{equation}
The above reduces in the nonrelativistic limit ($\xi>>1$) to the expression used earlier $\rho_{\nu}= m_{\nu}n_{\nu}$, where
\begin{equation}
n_{\nu}= \frac{T^3}{\pi^2}\int^{\infty}_0\frac{dyy^2}{e^y+1}=\frac{3\zeta(3)}{2\pi^2}T^3.
\end{equation}
For relativistic neutrinos ($\xi \rightarrow 0$), on the other hand, $\rho_{\nu}$ takes the familiar black body form
\begin{equation}
\rho_{\nu}=\frac{7\pi^2}{120}T^4.
\end{equation}

With $\rho_{\nu}$ given by the more general expression (12), the demand that $\rho_{dark}$ be stationary with respect to changes in $m_{\nu}$ implies that
\begin{equation}
\frac{\partial \rho_{dark}}{\partial m_{\nu}}= T^3\frac{\partial F}{\partial \xi} + \frac{\partial \rho_{dark~
energy}}{\partial m_{\nu}} =0.
\end{equation}

To derive the equation of state for the dark sector, let us examine the energy conservation equation, Eq. (8). Using
\begin{equation}
\rho_{{\rm dark}}=\rho_{\nu} +\rho_{\rm{dark ~energy}}
\end{equation}
and $p_{\rm{dark}}=\omega\rho_{\rm{dark}}$, Eq. (8) becomes
\begin{equation}
\dot{\rho}_{\rm{dark}}=- 3H\rho_{\rm{dark}}(1+ \omega).
\end{equation}
Now
\begin{equation}
\dot{\rho}_{\rm{dark}}= \dot{T}\frac{\partial \rho_{\rm{dark}}}{\partial T}=
-HT\frac{\partial \rho_{\rm{dark}}}{\partial T}.
\end{equation}
Hence, Eq. (18) reduces to
\begin{equation}
T\frac{\partial \rho_{\rm{dark}}}{\partial T}=
 3\rho_{\rm{dark}}(1+ \omega).
\end{equation}

The temperature dependence in $\rho_{\rm{dark}}$ is both explicit and through the dependence of the neutrino mass on temperature engendered by Eq. (16). One has
\begin{eqnarray} 
T\frac{\partial \rho_{\rm{dark}}}{\partial T}&=& 4\rho_{\nu} +T^5\frac{\partial \xi}{\partial T}\frac{\partial F}{\partial \xi}+ T\frac{\partial m_{\nu}}{\partial T}\frac{\partial \rho_{\rm{dark~energy}}}{\partial{m_{\nu}}}\\
&=&4\rho_{\nu} -T^4\xi\frac{\partial F}{\partial \xi} +T\frac{\partial m_{\nu}}{\partial T}[T^3\frac{\partial F}{\partial \xi}+\frac{\partial \rho_{\rm{dark~energy}}}{\partial{m_{\nu}}}].
\end{eqnarray}
However, the last term above vanishes as a consequence of Eq. (16). Thus, finally, defining
\begin{equation}
h(\xi)=\frac{\xi\frac{\partial F(\xi)}{\partial \xi}}{F(\xi)}
\end{equation}
one finds for the equation of state
\begin{equation}
\omega +1= \frac {\rho_{\nu}[4 -h(\xi)]}{3\rho_{dark}}.
\end{equation}
In the nonrelativistic limit ($\xi>>1$), it is easy to see that $h(\xi) \rightarrow 1$ and $\rho_{\nu} \rightarrow m_{\nu}n_{\nu}$. Thus, in this limit, Eq. (24) reduces to the FNW equation of state, Eq.(10).

The general equation of state (24) allows one to deduce an important result which is not obvious if one just examines its nonrelativistic limit, Eq. (10). For this purpose consider
\begin{equation}
(1 + \omega) \rho_{\rm{dark}}= \rho_{\rm{dark}} + p_{\rm{dark}}=\rho_{\nu} + p_{\nu}+\rho_{\rm{dark~energy}} + p_{\rm{dark~energy}}.
\end{equation}
Using the above, one can rewrite the equation of state as
\begin{equation}
  p_{\nu}+\rho_{\rm{dark~energy}} + p_{\rm{dark~energy}}= \frac{\rho_{\nu}}{3}[1-h(\xi)]= \frac{T^4}{3}[F(\xi)- \xi\frac{\partial F}{\partial \xi}].
\end{equation}
However,\begin{equation}
\frac{1}{3}[F(\xi)- \xi\frac{\partial F}{\partial \xi}]=\frac{1}{3\pi^2}\int^{\infty}_0\frac{dyy^4}{\sqrt{y^2+\xi^2}(e^y+
1)},
\end{equation}
and one recognizes that the RHS of Eq. (26) is just the canonical expression for the neutrino pressure: \newline $p_{\nu}=1/3<p^2/E>$. \cite{Weinberg} With this identification, we see from Eq. (26) that the FNW scenario requires that
\begin{equation}
p_{\rm{dark ~energy}}+\rho_{\rm{dark~energy}}=0.
\end{equation}
Thus, this scenario is only compatible with having a dark energy sector that
is described by a {\bf pure potential energy} term:
\begin{equation}
\rho_{\rm{dark~ energy}}(m_{\nu})= V(m_{\nu}).
\end{equation}
In essence, the dark energy in this scenario is equivalent to having a
cosmological constant, but one that {\bf{varies}} as a function of the
neutrino mass. If $m_{\nu}$ did not depend on temperature this would just correspond to the dark energy being given by a cosmological constant. However, here the cosmological constant {\bf{runs}}, since as the Universe expands the neutrino mass, and hence $V$, changes. \cite{SS}

The result given by Eq. (28), in effect, is perhaps not so surprising because in deriving this equation an implicit assumption was made regarding the temperature dependence of $\rho_{\rm{dark~energy}}$. Namely, what was assumed is that  $\rho_{\rm{dark~energy}}$ depended on temperature only through the temperature dependence of the neutrino mass. In general if  $\rho_{\rm{dark~energy}}$ had a kinetic term $K$, besides the potential term $V(m_\nu)$, this term can have an intrinsic temperature dependence: $K=K(T)$. Writing
\begin{equation}
 \rho_{\rm{dark~energy}}= K(T) +V(m_{\nu}(T)),
\end{equation}
Eq. (21) would acquire an additional contribution, $T\frac{\partial K(T)}{\partial T}$. This term gives a further contribution of $T\frac{\partial K(T)}{\partial T}/3\rho_{\rm{dark}}$ to the equation of state, Eq. (24), which, in turn, modifies Eq. (28) to
\begin{equation}
p_{\rm{dark ~energy}}+\rho_{\rm{dark~energy}}= \frac{T}{3}\frac{\partial K}{\partial T}.
\end{equation}

Because
\begin{equation}
p_{\rm{dark~energy}}= K(T) -V(m_{\nu}(T)),
\end{equation}
the FNW scenario implies that $K(T)$ obeys the differential equation
\begin{equation}
 K(T) =\frac{T}{6}\frac{\partial K}{\partial T},
\end{equation}
 whose solution is simply
\begin{equation}
 K(T) =K_o(\frac{T}{T_o})^6.
\end{equation}
That is, the kinetic energy term for dark energy is that of a free massless field. \footnote{This result is not unexpected. Consider, for example a quintessence model involving a scalar field $\phi(m_\nu)$. Then $K=\dot{\phi}^2/2$ and Eq. (33) reduces to $\ddot{\phi} + 3H\dot{\phi}=0$, which is indeed the equation of motion of a free scalar field. Even though $\rho_{\rm{dark~ energy}}$ has a potential term, the demand that $\rho_{\rm{dark}}$ be stationary under $m_\nu$ variations effectively forces $\partial V/\partial \phi +\partial \rho_{\rm{dark~ energy}}/\partial \phi$ to vanish also.} Although one can choose $K_o$ so as to make the kinetic energy contribution negligible in comparison to $V(m_{\nu})$ in the present epoch, in 
earlier times $K(T)$ totally dominates and distorts the evolution of the Universe. Hence, the FNW scenario is only consistent if there is no kinetic energy contribution ($K=0$) and the dark energy is a pure running cosmological constant.

\section{Two illustrative Examples}

We have seen in the preceding Section that the FNW scenario is characterized by two equations, Eqs (16) and (24). The first of these follow from the demand that $\rho_{\rm dark} $ be stationary with respect to $m_{\nu}$ variations. In view of Eq. (29), this stationarity requirement reduces to
\begin{equation}
T^3\frac{\partial F}{\partial \xi} + \frac{\partial V(m_{\nu})}{\partial m_{\nu}} =0.
\end{equation}
For any given potential $V$ characterizing the dark energy, the above equation determines the dependence of $m_{\nu}$ on the temperature $T$ or, equivalently, the neutrino number density $n_{\nu}$. The second fundamental equation for the scenario is the equation of state, which can be rewritten as
\begin{equation}
\omega +1= \frac {4-h(\xi)}{3[1+\frac{V(m_{\nu})}{T^4F(\xi)}]},
\end{equation}
For a given $V$, the evolution of $\omega$
with temperature follows from the above and the previously
determined dependence of $m_{\nu}$ on $T$.

I have studied two different potential models to illustrate the consequences of the FNW scenario. Because $\frac{\partial F}{\partial \xi}>0$, the slope of $V$ must be negative while, in general, $V$ is positive definite. This suggest two simple ansatze for the potential $V$. In the first of these, the potential has a power law dependence on $m_{\nu}$, $V \sim
m_{\nu}^{-\alpha}$. In the second ansatz, the dependence on $m_{\nu}$ is exponential, $V \sim  e^{\beta /m_{\nu}}$. In
both models the neutrino mass at high temperatures is much smaller
than its value now and the neutrinos become non-relativistic at
later times than if they had a fixed mass. 

 As long as neutrinos are relativistic,  their
contribution to the Universe's energy density is the standard one. In this limit, as we shall see, the neutrino contribution to
$\rho_{dark}$ dominates that of the potential term $V$. Thus,
at high temperatures, $\omega \simeq \omega_{\nu}= 1/3$. The value of
$\omega$ does not change much from this value until the neutrinos become
non-relativistic. At that point the dark energy potential begins to
become important and $\omega$ is rapidly driven to negative values. 
When $\omega$ approaches its value now, $\omega_o$, the dark sector energy
density dominates the Universe's energy density.

\subsection{ Power-law Potential}

 It is convenient to write the power-law potential as
\begin{equation}
V(m_{\nu})= - m^o_{\nu} n^o_{\nu} [\frac{\omega_o}{1 + \omega_o}](\frac{m_{\nu}}{ m^o_{\nu}})^{\frac{1 + \omega_o}{\omega_o}}.
\end{equation}
As we shall see, the constants $ m^o_{\nu}$, $ n^o_{\nu}$ and $\omega_o$ in the above are, respectively, the neutrino mass, the neutrino number density and the value of $\omega$ in the present epoch. The first two identifications follow by examining for this model Eq. (35), which determines the temperature dependence of the neutrino mass. For the potential (37), this equation becomes simply
\begin{equation}
 n^o_{\nu} (\frac{m_{\nu}}{ m^o_{\nu}})^{\frac{1 }{\omega_o}}=T^3 \frac{\partial F}{\partial \xi}= \frac {T^3}{\pi^2}\int^{\infty}_0\frac{dyy^2 \xi}{\sqrt{y^2+\xi^2}(e^y+1)}.
\end{equation}
As $T$ approaches the temperature of neutrinos now $T_o= 1.69 \times 10^{-4} $ eV, provided $\xi_o =m^o_{\nu}/T_o>>1$ (which turns out to be the case), the RHS of Eq. (38) approaches $n^o_{\nu}$ and, therefore, in this limit $ m_{\nu} \to m^o_{\nu}$.

Using Eqs (37) and (38) one easily see that
\begin{equation}
\frac{V}{T^4F(\xi)}= -\frac{\omega_o}{1 + \omega_o}\frac{\xi \frac{\partial F}{\partial \xi}}{F(\xi)}=-\frac{\omega_o}{1 + \omega_o}h(\xi).
\end{equation}
Thus the equation of state for the power-law potential reduces simply to
\begin{equation}
\omega +1= \frac {4-h(\xi)}{3[1-\frac{\omega_o}{1 + \omega_o}h(\xi)]}.
\end{equation}
Since for non-relativistic neutrinos $\xi>>1$ and $h(\xi) \to 1$, as anticipated, one sees that in the present epoch $ \omega \to \omega_o$. Therefore, in view of Eq. (10), one has
\begin{equation}
\omega_0 +1= \frac {m^o_{\nu}n^o_{\nu}}{\rho^o_{\rm{dark}}}= \frac{m^o_{\nu}n^0_{\nu}}{
m^0_{\nu}n^o_{\nu}+V(m^0_{\nu})}.
\end{equation}

From the above equation one can determine the present neutrino mass $m^o_{\nu}$ from a knowledge of the cosmological parameters $\omega_o$ and $\rho^o_{\rm{dark}}$. In what follows, we shall take $\omega_o=-0.9$ and $\rho^o_{\rm{dark}}=0.7 \rho_c$. Here $\rho_c$ is the critical density needed to close the Universe: $\rho_c= 3H^2_o/8\pi G_N =( 2.73 \pm 0.36) \times 10^{-11}$ e$V^4$ and the values assumed for $\omega_o$ and $\rho^o_{\rm{dark}}$ are good central values inferred from cosmological data. \cite{Omega} It follows from Eq. (41) then that
\begin{equation}
m^o_{\nu}n^o_{\nu}=0.1\rho^o_{\rm{dark}};~~~ 
V(m^0_{\nu})=0.9\rho^o_{\rm{dark}}.
\end{equation}
Using the central value for $\rho_c$, one finds that $V(m^0_{\nu})=2.46 \times 10^{-11}$ e$V^4$ and $m^o_{\nu}n^o_{\nu}=2.73 \times 10^{-12}$ e$V^4$. Since $n^o_{\nu}= 8.83 \times 10^{-13}$ e$V^3$, with these assumptions the neutrino mass in the present epoch has the value $m^o_{\nu}= 3.09$ eV.\footnote{ As Fardon {\it{ et al}} \cite {FNW} point out, such values for $m^o_{\nu}$ are not necessarily in contradiction with terrestrial limits on neutrino masses since the neutrino mass meausured on earth may well differ from $m^o_{\nu}$, if there is an overdensity in the local group due to neutrino clustering. \cite {RW}}

Having fixed the parameters $m^o_{\nu}$ and $\omega_o$, it is now straightforward to solve Eqs (38) and (40) numerically. Using Eq. (14) and defining
\begin{equation}
g(\xi)= \frac{\int^{\infty}_0\frac{dyy^2 \xi}{\sqrt{y^2+\xi^2}(e^y+1)}}{\int^{\infty}_0\frac{dyy^2 }{(e^y+1)}},
\end{equation}
Eq. (38) becomes
\begin{equation}
m_{\nu}= m^o_{\nu}[ (\frac{T}{T_o})^3 g(\xi)]^{\omega_o}.
\end{equation}
In the non-relativistic limit ($\xi>>1$) $g(\xi) \to 1$ and Eq. (44) gives
\begin{equation}
m_{\nu}\simeq m^o_{\nu} (\frac{T}{T_o})^{3 \omega_o}= \frac{2.02 \times 10^{-10} \rm{eV}}{[T(\rm{eV})]^{2.7}} ~~~[\rm{NR ~regime}],
\end{equation}
in agreement with Eq. (11).

The transition between the non-relativistic regime and the relativistic regime occurs in the neighborhood of $ \xi= m_{\nu}/T=1$ at a temperature
\begin{equation}
T^*= m_{\nu}(T^*) =\frac{(m^o_{\nu})^{1/3.7}(T_o)^{2.7/3.7}}{[g(1)]^{0.9}}= 3.06 \times 10^{-3} \rm{eV}
\end{equation}
which, as anticipated, is much below the transition temperature for {\bf fixed mass} neutrinos, $ T^*_{\rm fixed}= 3.09$ eV.

In the relativistic regime ($\xi \to 0$), on the other hand, since $ g(\xi) \to 0.456 \xi$ one finds that $m_{\nu}$ decreases nearly linearly with temperature:
\begin{equation}
m_{\nu}\simeq \frac{1.12 \times 10^{-5} \rm{eV}}{[T(\rm{eV})]^{0.95} }~~~[\rm{Relativistic ~regime}].
\end{equation}
The full behavior of $m_{\nu}$ as a function of temperature for the power-law potential, given by Eq. (44), is plotted in Fig. 1.

\begin{figure}[t]
\centering
\hspace*{-5.5mm}
\leavevmode\epsfysize=6cm \epsfbox{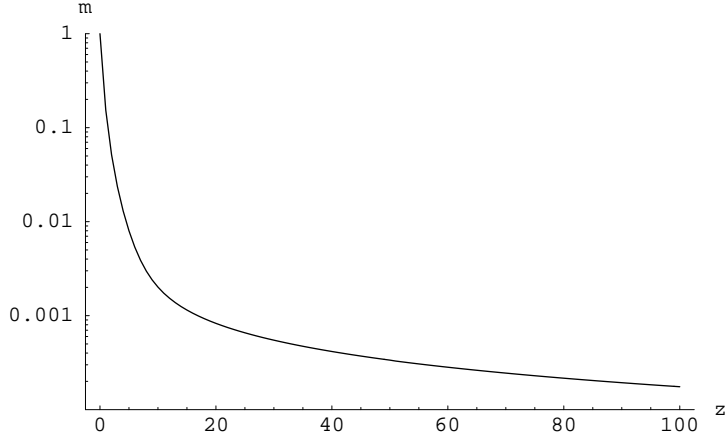}\\[3mm]
\caption[Fig. 1]{\label{Fig. 1} Plot of the scaled mass $m=m_{\nu}/m^o_{\nu}$ versus the redshift $ z=(T/T_o-1)$ for the power-law potential.}
\end{figure}

Once $m_{\nu}(T)$ is known, it is straightforward to compute from Eq. (40) the behaviour of the equation of state with temperature. As noted earlier, in the non-relativistic limit ($\xi>>1$) $ \omega \to \omega_o= -0.9$. At the transition temperature $T^*$, since $ h(1)= 0.111 $, one finds $\omega(T^*)=-0.351$. Finally, in the relativistic regime ($\xi \to 0$), since $h(\xi) \to 0$, $\omega \to 1/3$. Fig. 2 shows the full evolution of the equation of state with temperature, or equivalently with the redshift $ z= T/T_o -1$, for the power-law potential model.

\begin{figure}[t]
\centering
\hspace*{-5.5mm}
\leavevmode\epsfysize=6cm \epsfbox{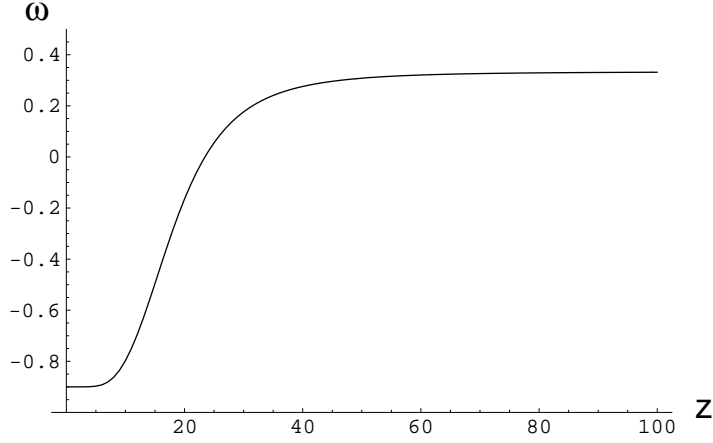}\\[3mm]
\caption[Fig. 2]{\label{Fig. 2} Plot of the equation of state parameter $\omega$ versus the redshift $ z=(T/T_o-1)$ for the power-law potential.}
\end{figure}

\subsection{Exponential Potential}

As a second illustrative example it is useful to consider an exponential form for the potential $V(m_{\nu})$. A convenient way to write this potential is
\begin{equation}
V(m_{\nu})= - m^o_{\nu} n^o_{\nu} [\frac{\omega_o}{1 + \omega_o}]e^{-\frac{1 + \omega_o}{\omega_o}
[(\frac{m^o_{\nu}}{ m_{\nu}})-1]}
\end{equation}
Here the parameters  $ m^o_{\nu}$, $ n^o_{\nu}$ and $\omega_o$ have precisely the same meaning as in the previous subsection and take the numerical values assumed there.

For the potential (48), Eq. (35) which determines the temperature dependence of the neutrino mass becomes
\begin{equation}
(\frac{ m^o_{\nu}}{ m_{\nu}})^2 e^{-\frac{1 + \omega_o}{\omega_o}
[(\frac{m^o_{\nu}}{ m_{\nu}})-1]}=(\frac{T}{T_o})^3g(\xi)
\end{equation}
In the non-relativistic limit $(\xi >>1)$, since $g(\xi) \to 1$, in leading approximation\begin{equation}
m_{\nu} \simeq m^o_{\nu}(\frac{T_o}{T})^{3/2}=
\frac{6.79 \times 10^{-6} {\rm eV}}{ [T({\rm eV})]^{1.5}}~~~[\rm{NR ~regime}].
\end{equation}
Note that for this potential the temperature dependence of $m_{\nu}$ in the non-relativistic limit does not follow Eq. (11) since, as we shall see below, even in this limit $\omega$ changes rather rapidly with temperature. 

For this potential, the transition between the non-relativistic  and the relativistic regime ($ \xi= m_{\nu}/T=1$)  occurs at a temperature $T^*$  which is the solution of the trascendental equation
\begin{equation}
\frac{e^{0.343/T^*({\rm eV})}}{T^{*5}({\rm eV})}= 9.01\times 10^9.
\end{equation}
Numerically, one finds that $T^*= 4.57 \times 10^{-2} {\rm eV}$ which is about a factor of 10 greater than the transition temperature for the power-law potential, but still much below the transition temperature for fixed mass neutrinos, $ T^*_{\rm fixed}= 3.09$ eV.

In the relativistic regime $(\xi \to 0)$, using that $g(\xi) \to 0.456 \xi$,  $m_{\nu}$ is given by the solution of the trascendental equation
\begin{equation}
m_{\nu}(\rm{eV})=\frac{0.343}{23.126+ 3~\rm{ln}~m_{\nu}(\rm {eV}) +2~\rm{ln}~T(\rm {eV})} ~~~[\rm{Relativistic~regime}].
\end{equation}
One sees from the above that the neutrino mass decreases only logarithmically with temperature, in contrast to the nearly linear decrease with temperature for $m_{\nu}$ in the power-law potential. For instance, for $T=1$ eV, $m_{\nu}= 0.028$ eV. The full behavior of $m_{\nu}$ as a function of temperature for the exponential potential, obtained by numerically solving Eq. (44), is  displayed in Fig. 3.

\begin{figure}[t]
\centering
\hspace*{-5.5mm}
\leavevmode\epsfysize=6cm \epsfbox{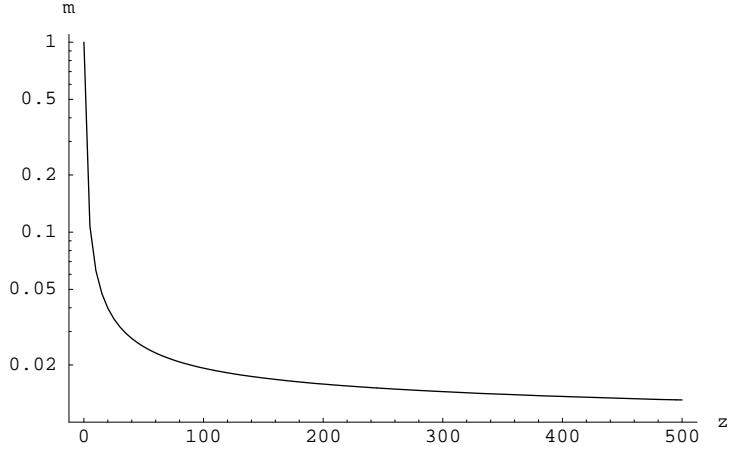}\\[3mm]
\caption[Fig. 3]{\label{Fig. 3} Plot of the scaled mass $m=m_{\nu}/m^o_{\nu}$ versus the redshift $ z=(T/T_o-1)$ for the exponential potential.}
\end{figure}

Using Eqs. (48) and (49) one has for the exponential potential
\begin{equation}
\frac{V}{T^4F(\xi)}= -\frac{\omega_o}{1 + \omega_o}(\frac{m_{\nu}}{m^o_{\nu}})\frac{\xi \frac{\partial F}{\partial \xi}}{F(\xi)}=-\frac{\omega_o}{1 + \omega_o}(\frac{m_{\nu}}{m^o_{\nu}})h(\xi).
\end{equation}
Hence, for this potential, the equation of state (36) is simply
\begin{equation}
\omega +1= \frac {4-h(\xi)}{3[1-\frac{\omega_o}{1 + \omega_o}(\frac{m_{\nu}}{m^o_{\nu}})h(\xi)]}.
\end{equation}
This is similar to the equation of state for the power-law potential, Eq. (40), apart from the extra factor of $m_{\nu}/m^o_{\nu}$ in the denominator. However, because of this additional factor, even in the non-relativistic limit where $ \xi>>1$ and $ h(\xi) \to 1$, the equation of state has an explicit temperature dependence which follows from Eq. (50). That is, in this limit, Eq. (54) reduces to
\begin{equation}
\omega = -\frac{9(\frac{T_o}{T)})^{3/2}}{1+9(\frac{T_o}{T})^{3/2}}~~~~[\rm {NR~limit}].
\end{equation}
Hence, for example, already at $T=2T_o=3.38 \times 10^{-4}$ eV one finds $\omega(2T_o)=-0.777$, which is quite different from its value in the present epoch $\omega(T_o)=\omega_o=-0.9$. At the transition temperature $T^*=4.57 \times 10^{-2}$ eV, since $h(1)=0.111$, Eq. (54) gives $\omega(T^*)=0.287$. This value is already close to the asymptotic value which $\omega$ attains in the relativistic limit, $\omega \to 1/3$. Fig. 4 shows the full evolution of the equation of state with temperature for the exponential potential.

\begin{figure}[t]
\centering
\hspace*{-5.5mm}
\leavevmode\epsfysize=6cm \epsfbox{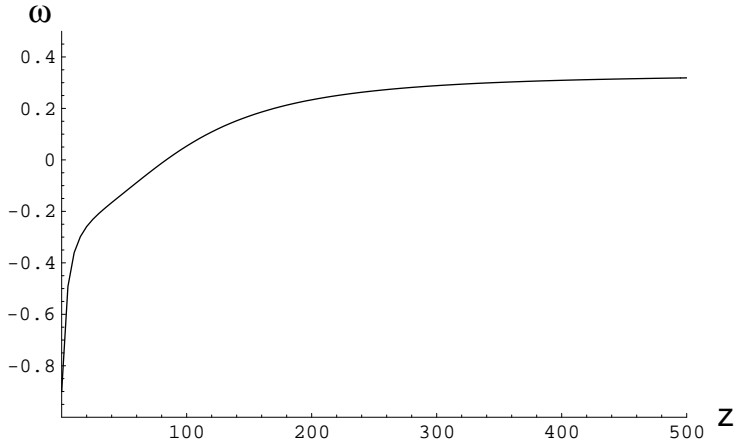}\\[3mm]
\caption[Fig. 4]{\label{Fig. 4} Plot of the equation of state parameter $\omega$ versus the redshift $ z=(T/T_o-1)$ for the exponential potential}
\end{figure}

\subsection {Lessons Learned from the Model Examples}

It is useful to examine the evolution of the energy density of the Universe in the FNW scenario. By assumption
\begin{equation}
\rho=\rho_{\rm{matter}} +\rho_{\rm{dark}},
\end{equation}
with
\begin{equation}
\rho_{{\rm dark}}=\rho_{\nu} +V_{\rm{dark~ energy}}=T^4F(\xi(T))+V(m_{\nu}(T)).
\end{equation}
In our discussion above, we assumed that in the present era the two components of $\rho$ in Eq. (56) accounted, respectively for 30\% and 70\% of $\rho_c$. Because $p_{\rm{matter}}=0$, for $T>T_o$ the matter energy density scales as $(T/T_o)^3$:
\begin{equation}
\rho_{\rm{matter}}=\rho^o_{\rm{matter}}(\frac{T}{T_o})^3= 0.3\rho_c (\frac{T}{T_o})^3.
\end{equation}
At high temperatures, the density of the dark sector is dominated by $\rho_{\nu}$, since this energy density in the relativistic regime grows like $T^4$. As we shall see below, this is much faster than the growth with temperature of the potential $V(m_{\nu}(T))$.

In the two models discussed we specifically assumed that $\rho_{\nu}(T_0)=0.07\rho_c$ and $V(m^0_{\nu})=0.63\rho_c$ [cf Eq. (42)]. Using this input we can rewrite $\rho_{\nu}(T)$ as 
\begin{equation}
\rho_{\nu}(T)= 0.07 \rho_c \frac{7\pi^2}{180\zeta(3)}(\frac{T_o}{m^o_{\nu}})k(\xi(T))(\frac{T}{T_o})^4=1.21 \times 10^{-5} \rho_c k(\xi(T))(\frac{T}{T_o})^4.
\end{equation}
Here
\begin{equation}
k(\xi)= \frac{120}{7\pi^2}F(\xi),
\end{equation}
and this function is normalized so that in the relativistic limit, $\xi \to 0$,
$k(\xi) \to 1$. We see from Eq. (59) that the neutrino energy density is equal to the matter energy density at a temperature $T_{\rm{eq}}= 2.48 \times 10^{4} T_o=4.19$ eV. \footnote{ In our discussion we have ignored altogether the photon contribution to the energy density of the Universe, since it gives a negligible contribution to the energy density of the Universe in the present epoch. However, at high temperatures its contribution is comparable to that of the neutrino energy density: $\rho_\gamma=(8/7) \rho_\nu= \pi^2T^4/15$.}

The temperature dependence of the dark energy contribution depends in detail on the potential $V(m_{\nu})$ one assumes, and the concomitant dependence of the neutrino mass on temperature. For the power-law potential of Eq. (37), using Eq. (44) one has
\begin{equation}
V_{\rm {p}}(T)=0.63 \rho_c[(\frac{T}{T_o})^3g(\xi)]^{0.1}.
\end{equation}
In the relativistic regime, $\xi \to 0$, one has using Eq. (47)
\begin{equation}
g(\xi) \to 0.456 \frac{m_{\nu}}{T}=\frac{ 5.11 \times 10^{-6}}{[T(\rm{ eV})]^{1.95}}.
\end{equation}
Thus, in this limit, one finds for the power-law potential
\begin{equation}
V_{\rm {p}}(T)=2.52 \rho_c[T(\rm{eV})]^{0.105}~~~~[\rm{Relativistic~regime}]
\end{equation}
From the above, it is clear that $V_{\rm{p}}$ is nearly temperature independent and is totally negligible compared to $\rho_{\nu}$ for $T>>T_o$. The potential (61) only begins to dominate $\rho_{\rm{dark}}$ (and $\rho$) for temperatures very near to $T_o$. Fig. 5 shows the temperature behaviour of the various components of the energy density of the Universe for the power-law potential model.

\begin{figure}[t]
\centering
\hspace*{-5.5mm}
\leavevmode\epsfysize=6cm \epsfbox{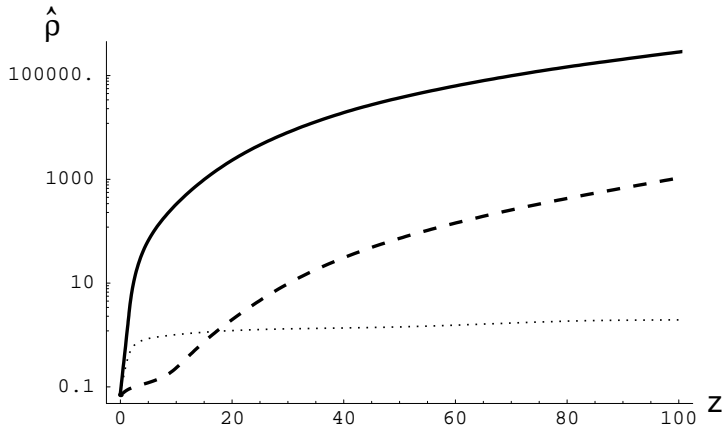}\\[3mm]
\caption[Fig. 5]{\label{Fig. 5} Plot of the various components of the Universe's energy density in units of $\rho_c$, $\hat{\rho}=\rho/\rho_c$, versus the redshift $ z=(T/T_o-1)$ for the power-law potential. In the Figure $\hat{\rho}_{\rm{matter}}$ is given by the solid line; $\hat{\rho}_{\nu} $ is given by the dashed line; and $\hat{V}_{\rm{p}}$ is given by the dotted line. }
\end{figure}

Similar considerations apply for the case of the exponential potential (48). Using Eq. (49) one can write
\begin{equation}
V_{\rm {e}}(T)=0.63\rho_c(\frac{m_{\nu}}{m^o_{\nu}})^2(\frac{T}{T_o})^3 g(\xi).
\end{equation}
In the relativistic regime, $\xi \to 0$, since $g(\xi) \to 0.456 \xi$ for this potential one finds
\begin{equation}
V_{\rm {e}}(T)=0.29\rho_c\frac{m_{\nu}^3T^2}{m^{o2}_{\nu}T_o^3} ~~~~~[\rm{Relativistic~ regime}].
\end{equation}
Because in this regime the neutrino mass is only slowly varying with temperature [cf Eq. (52)], one sees from Eq. (65) that the exponential potential grows essentially quadratically with temperature, $V_{\rm{e}} \sim T^2$. Numerically, for example at $T=1$ eV, $V_{\rm{e}}(1~{\rm{eV}})=4.33 \times 10^4 \rho_c$. This energy density is much greater than the corresponding one for the power-law potential, $V_{\rm{p}}(1~\rm{eV})= 2.52 \rho_c$, but is still much below the value of the neutrino energy density at this temperature, $\rho_{\nu}(1~\rm{eV})= 1.48 \times 10^{10} \rho_c$. Fig. 6 shows the behaviour of the various components of the energy density of the Universe as a function of temperature for the exponential potential. One sees from Figs. 5 and 6 that the consmological constant contribution to the Universe's energy density is subdominant at high temperatures. Thus, for example, the dark energy in these models is totally negligible at the time of Nucleosynthesis and the FNW scenario is not constrained by what happens in this epoch.

\begin{figure}[t]
\centering
\hspace*{-5.5mm}
\leavevmode\epsfysize=6cm \epsfbox{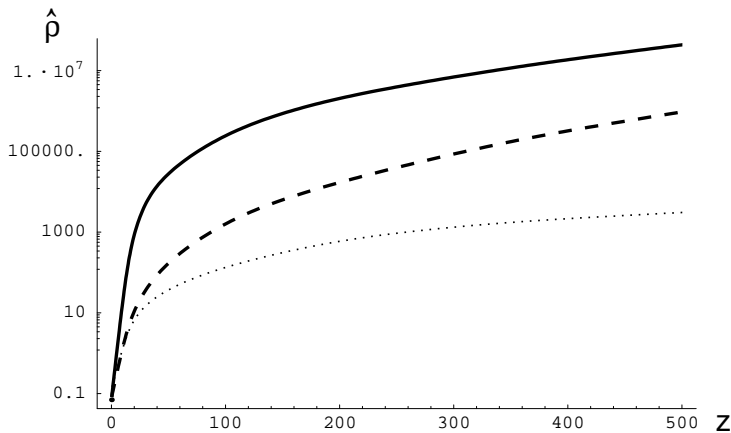}\\[3mm]
\caption[Fig. 6]{\label{Fig. 6} Plot of the various components of the Universe's energy density in units of $\rho_c$, $\hat{\rho}=\rho/\rho_c$, versus the redshift $ z=(T/T_o-1)$ for the exponential potential. In the Figure $\hat{\rho}_{\rm{matter}}$ is given by the solid line; $\hat{\rho}_{\nu} $ is given by the dashed line; and $\hat{V}_{\rm{e}}$ is given by the dotted line. }
\end{figure}

\section{Discussion}

There are many aspects of the FNW scenario that are very intriguing. However, there are also features which are unsatisfactory. I think it is very interesting to think of the dark energy in the Universe as a cosmological constant which evolves as the Universe evolves. However, it is difficult to believe that the only relevant parameter for this evolution is just the neutrino mass scale. Although neutrino masses can be effectively connected to an $SU(2) \times U(1)$ invariant scale associated with heavy right-handed neutrinos, \cite{seesaw} \cite {seesaww} it is hard to countenance that only this scale should affect the evolution of the cosmological constant.

A second unsatisfactory aspect of the FNW scenario is related to the intrinsic scale of the dark energy. Roughly speaking, this scale is "put in by hand" by fixing the value of the potential $V$ in the present epoch: $V(m^o_{\nu}) \simeq T^3_om^o_{\nu} \simeq E^4_o$. Different potential models then dictate the precise temperature dependence of $V$, so that as the Universe evolves the cosmological constant varies as
\begin{equation}
V\simeq E^4_o f(T/T_o).
\end{equation}
In the two models we discussed, roughly, $f_{\rm{p}} \sim \rm{constant}$ and $f_{\rm{e}} \sim (T/T_o)^2$. Naively, one might expect that vacuum energies generated  at a temperature $T$ should be of order $T^4$. Thus, an equation like (66) with an intrinsic scale $E_o$ seems quite mysterious.

A third unconfortable feature of the FNW scenario is that, in essence, no dynamics can be associated with the dark energy sector, save for the "running" of the cosmological constant as a function of the variation of the neutrino mass with temperature.  At first sight, it would seem more likely that one should attribute the variation of the cosmological constant to the presence of some dynamical field. In this respect, a more appealing scenario could be one which I explored long ago with Sola and Wetterich, where the cosmological constant changed in response to changes in a dynamical field -- the cosmon. \cite{cosmon}

In the cosmon model, the presence of a spontaneously broken dilatation symmetry drove the anomalous trace of the energy momentum tensor $\theta^{\mu}_{\mu}$ to zero, by a mechanism analogous to how a $U(1)_{PQ}$ symmetry \cite{PQ} drives the strong CP parameter $\theta$ to zero. In \cite{cosmon}, we imagined that the full trace of the energy momentum tensor $T^{\mu}_{\mu}$ was proportional to the anomalous trace of this tensor so that, effectively, the cosmon mechanism caused the cosmological constant to vanish. However, we could never convincingly establish this proportionality. On the other hand,  if  $T^{\mu}_{\mu}$ were {\bf not} to be proportional to $\theta^{\mu}_{\mu}$, the cosmon mechanism could effectively determine the cosmological constant as a result of a minimization principle.

If $S$ is the cosmon field which translates under scale transformations, $S \to S +\alpha M$ (with $M$ being the scale of the spontaneous breaking of the dilatational symmetry -- presumably $M \sim M_P$), then one can imagine the cosmological constant ensuing as a solution of the equation:
\begin{equation}
M\frac{\partial \rho}{\partial S}|_{S=S_o}=0.
\end{equation}
The above equation is precisely the type of equation considered in the FNW scenario, with the dynamical field $S$ being replaced by the neutrino mass $m_{\nu}$. More precisely, $m_{\nu}\to m_{\nu}e^{S/M}$ and $S_o=M$.

 With an equation like (67) one needs to worry how other mass scales, besides neutrino masses, depend on the dynamical field $S$. Furthermore, if $S$ is relatively light, one has also to consider how its presence can give rise to violations of the equivalence principle. \cite{cosmon} \cite{chamaleon} I will not pursue these matters further here.  At any rate, given the dearth of realistic
particle physics explanations for the observed dark energy component of
the Universe, it appears worthwhile to continue to explore these and other speculations stimulated by the FNW scenario.

I would like to thank S. Bludman, W. Buchm$\ddot{\rm u}$ller, G. Gelmini, A. Kusenko, A. Ringwald, T. Yanagida and X. M. Zhang for very helpful discussions. This work was supported in part by the Department of Energy under Contract No. FG03-91ER40662, Task C. I am grateful to W. Buchm$\ddot{\rm u}$ller for his hospitality at DESY, where some of this work was carried out.


\begin{thebibliography}{99}

 
\bibitem{FNW}  R. Fardon, A. E. Nelson and N. Weiner, J. Cosmol. Astropart. Phys. {\bf 10}, 005 (2004)

\bibitem{tracking} P. J. Steinhardt, L.Wang, and I. Zlatev, Phys. Rev.  {\bf D59}, 123504 (1999); see also \cite{Q} \cite{q}

\bibitem{variable} J. A. Casas, J. Garcia- Bellido, and M. Quiros, Class. Quant. Grav. {\bf 9}, 1371 (1992)

\bibitem{JGB} J. Garcia- Bellido, Int. J. Mod. Phys. {\bf D2}, 85 (1993)

\bibitem{PQH} P. Q. Hung, hep-ph/0010126

\bibitem{Q} C. Wetterich, Nucl. Phys. {\bf B302}, 668 (1988)

\bibitem{q} B. Ratra and P. J. E. Peebles, Phys. Rev. {\bf D37}, 3406 (1988)

\bibitem{qQ} R. R. Caldwell, R. Dave, and P. J. Steinhardt, Phys. Rev. Lett. {\bf 80}, 1582,(1998)

\bibitem{Omega} A. G. Riess {\it et al}, Astrophys. J. {\bf 607}, 665 (2004) 
        
\bibitem{Weinberg} S. Weinberg, {\bf Gravitation and Cosmology} ( John Wiley and Sons, New York, 1972)

\bibitem{SS} For earlier work on a running cosmological constant, see, I. L. Shapiro and J. Sola, Phys. Lett. {\bf B475}, 236 (2000)

\bibitem{RW} A. Ringwald and Y. Y. Y. Wong,  J. Cosmol. Astropart. Phys. {\bf 12}, 005 (2004)

 
\bibitem{seesaw} T. Yanagida, in Proceedings of the Workshop on the Unified
Theories and Baryon Number in the Universe, Tsukuba, Japan 1979, eds.
O. Sawada and A. Sugamoto, KEK Report No. 79-18 

\bibitem {seesaww} M. Gell-Mann, P. Ramond
and R. Slansky in {\bf Supergravity}, Proceedings of the Workshop at
Stony Brook, NY, 1979, eds. P. van Nieuwenhuizen and D. Freedman
(North-Holland, Amsterdam, 1979).

\bibitem{cosmon} J. Sola, R. D. Peccei, and C. Wetterich, Phys. Lett. {\bf B195}, 183 (1987)

\bibitem{PQ} R. D. Peccei and H. R. Quinn, Phys. Rev. Lett. {\bf 38}, 1440 (1977); Phys. Rev. {\bf 16}, 1791 (1977)

\bibitem{chamaleon} J. Khoury and A. Weltman, Phys. Rev. {\bf D69}, 044026 (2004)

\end{thebibliography}
\end{document}